\documentclass[conference]{IEEEtran}
\usepackage{spconf,amsmath,graphicx}

\usepackage{amsmath}
\usepackage{amsthm}
\usepackage{amssymb}
\usepackage{amsfonts}
\usepackage{xcolor}
\usepackage{mathtools}
\usepackage{verbatim}
\usepackage{multicol}
\usepackage{multirow}
\usepackage{mathabx}
\usepackage{graphicx}
\usepackage{booktabs}
\usepackage{float}
\usepackage{listings}
\usepackage{array}
\usepackage{etoolbox}
\usepackage{setspace}
\usepackage[hidelinks]{hyperref}
\usepackage{tikz}
\usepackage{textcomp}

\newcommand{\N}{{\mathbb{N}}}

\newcommand{\Z}{{\mathbb{Z}}}

\newcommand{\ba}{\begin{array}}
\newcommand{\ea}{\end{array}}

\newcommand\blfootnote[1]{%
  \begingroup
  \renewcommand\thefootnote{}\footnote{#1}%
  \addtocounter{footnote}{-1}%
  \endgroup
}

\let\sumnonlimits\sum
\renewcommand{\sum}{\sumnonlimits\limits}

\title{Hyperspectral reconstruction of skin through fusion of scattering transform features}

\twoauthors
  {Wojciech Czaja, Jeremiah Emidih, Brandon Kolstoe\sthanks{Corresponding Author: bkolstoe@umd.edu}}
	{University of Maryland\\
	Department of Mathematics\\
	College Park, Maryland, United States}
  {Richard G. Spencer}
	{NIH, National Institute on Aging\\
	Laboratory of Clinical Investigation\\
	Baltimore, Maryland, United States}
\newcommand\copyrighttext{%
  \footnotesize \copyright 2024 IEEE.  Personal use of this material is permitted. Permission from IEEE must be obtained for all other uses, in any current or future media, including reprinting/republishing this material for advertising or promotional purposes, creating new collective works, for resale or redistribution to servers or lists, or reuse of any copyrighted component of this work in other works.
}

\renewcommand\copyrightnotice{%
\begin{tikzpicture}[remember picture,overlay]
\node[anchor=south,yshift=10pt] at (current page.south) {\fbox{\parbox{\dimexpr\textwidth-\fboxsep-\fboxrule\relax}{\copyrighttext}}};
\end{tikzpicture}%
}

\begin{document}
%
\maketitle

\copyrightnotice
%
\begin{abstract}

\noindent Hyperspectral imagery (HSI) is an established technique with an array of applications, but its use is limited due to both practical and technical issues associated with spectral devices. The goal of the ICASSP 2024 `Hyper-Skin' Challenge is to extract skin HSI from matching RGB images and an infrared band. To address this problem we propose a model using features of the scattering transform - a type of convolutional neural network with predefined filters. Our model matches and inverts those features, rather than the pixel values, reducing the complexity of matching while grouping similar features together, resulting in an improved learning process. \blfootnote{\noindent This work was funded in part by the Intramural Research Program of the National Institute on Aging of the NIH.}

\end{abstract}
\section{Introduction}

Hyperspectral imagery (HSI), consisting of a set of 2D images representing different wavelengths, finds numerous applications in a wide range of fields \cite{HSIReview}. In spite of its value, in many applications there are practical limitations for implementing this technology. The goal of the ICASSP 2024 Grand Challenge on Hyperspectral Skin Vision \cite{hyperskin2024overview} is therefore to reconstruct HSI data from multispectral data (henceforth referred to as MSI) consisting of standard RGB images and one near-infrared band, with applications to skin analysis \cite{ng2023hyperskin}. We propose a model based on  matching and inverting scattering transform \cite{MallatOrig} features to address this challenge. 



\section{Methods}

\noindent{\bf 2.1. Overview.}
\vskip0.1cm
\noindent Our model takes advantage of representing and matching data in the feature space \cite{fusion}, instead of its original representation. Features are extracted by the scattering transform \cite{MallatOrig}, which groups information about size, location, and direction into scattering coefficients. Then, a transformation is learned by matching MSI scattering coefficients to those of the training HSI data. A preimage is then applied to obtain an HSI representation. Finally, a multi-image superresolution (MISR) network more closely aligns the skin values of the preimage. 

\vskip0.2cm

\noindent{\bf 2.2. Scattering Transform.}
\vskip0.1cm
\noindent The scattering transform uses a convolutional neural network (CNN) structure but with predefined wavelet filters \cite{MallatOrig}. Let $\psi$ be the mother wavelet, fix $L\in \N$, and for $j,q\in \Z$ denote $\psi_{j,q}(u)=2^{-2j}\psi(2^{-l}r_{\theta}u)$, 
where $\theta = q\pi /L$ and $r_\theta$ is the corresponding rotation. Also, let $\phi_J(u) = 2^{-2J}\phi(2^{-J}u)$ where $\phi$ is a low-pass filter. Then, the 2-layer scattering transform of an image $x$ is given by: 
\vskip-0.6 cm
\begin{equation*}
\{x*\phi_J,|x*\psi_{j,q}|*\phi_J,||x*\psi_{j,q}|*\psi_{j',q'}|*\phi_J\}_{\substack{1\leq j<j'\leq J\\ 1\leq q,q' \leq L}}
\end{equation*}
\vskip-0.2cm
\noindent where $*$ denotes (periodic) convolution. 
\vskip0.2cm

\noindent{\bf 2.3. Matching Networks.}
\vskip0.1cm
\noindent 
Due to computational constraints, two different matching networks are used. One network matches the scattering coefficients of MSI images to the scattering coefficients of even channels of HSI images, while the other network matches them to the coefficients of the odd channels. This was done to achieve compatible interpolations of both of these sets of HSI channels, rather than reconstructing results over essentially non-overlapping spectral bands. Each matching network consists of 2 linear layers. ReLU is used for the activation function of the first layer and $\tanh$ for the second. 

\vskip0.2cm

\noindent{\bf 2.4. Inverse Networks.}
\vskip0.1cm
\noindent 
As above, one inverse network is trained for scattering coefficients of even channels and another for odd channels. We use an inverse similar to the one proposed in \cite{ScatGAN}. Two convolutional blocks, each consisting of upsampling by a factor of 2, a convolutional layer, batch normalization, and ReLU activation, are first applied. This is followed by one more convolutional layer, batch normalization, and $\tanh$ activation. 
\vskip0.2cm

\noindent{\bf 2.5. Multi-image Superresolution Network.}
\vskip0.1cm
\noindent Some correlations are lost between adjoining HSI channels due to their separation above. This final network is applied to improve alignment of the predicted even and odd channel images. First, a mask is applied to the predicted HSI images which removes (most) non-skin features. Then, the network is applied to the skin spectra (i.e. the vectors of pixels from each HSI channel which have the same coordinates). The general network architecture is identical to the matching network. 




\section{Implementation and Results}

\noindent{\bf 3.1. Dataset.} 
\vskip0.1cm
\noindent We use the data in \cite{ng2023hyperskin}, as detailed in \cite{hyperskin2024overview}. Note that the HSI data is divided into the visual range (VIS), from 400 to 700 nm in 10 nm increments, and the near-infrared range (NIR), from 700 to 1000 nm in 10 nm increments, with 700 nm in common. We label the channels in VIS from 0 to 31 and NIR from 32 to 61, so 31 and 32 are the same channel. Even channels refer to the even labels, and similarly for odd channels. 
\vskip0.2cm



\noindent{\bf 3.2. Implementation Details.}
\vskip0.1cm
\noindent 
To compute the scattering transform, we use the package Kymatio \cite{JMLR:v21:19-047}. Here, $\psi$ is a Morlet wavelet and $\phi$ is a Gaussian. Additionally, the output is downsampled by a factor of $2^J$. For our purposes, we choose $J=2$ and $L=4$, and we take the scattering transform of each channel of our images. 

All network architectures were implemented using Pytorch, using the Adam optimizer with learning rate 0.001. The matching networks were trained for 100 epochs each using $L^2$ loss. The inverse networks were trained for 150 epochs each using $L^1$ loss. The MISR network was trained for either 30 or 60 epochs with $L^2$ loss.

Three models were tested. Model 1 doesn't use the MISR network, whereas Models 2 and 3 use the network trained for 30 and 60 epochs, respectively (in order to compare between different training times). Since the outputs of the models have one channel in common (700 nm), the final output for that channel is obtained by averaging these\footnote{Source codes for our models are available at \url{https://github.com/BrandonKolstoe/Hyperskin_Scattering}.}.

\vskip0.2cm

\noindent{\bf 3.3. Results.}
\vskip0.1cm
\noindent 
Table 1 summarizes the results of our three models and the baseline Mask-guided Spectral-Wise Transformer (MST++) method \cite{baseline}. Models were compared using the average of the spectral angle mapper (SAM) scores \cite{ng2023hyperskin} of the reconstructed skin values of the testing set; a lower SAM score corresponds to better performance. Our models perform comparably to the baseline model, with the third model performing best. 

\begin{table}
\caption{Average SAM scores of skin reconstructions of the MST++ baseline and our proposed models.}
\begin{center}
    \begin{tabular}{cc}
        \hline
        Model & Average SAM score\\
        \hline
        MST++ (baseline) & $0.1182 \pm 0.0200$  \\
        Model 1 (no MISR) & $0.1201 \pm 0.0116$  \\
        Model 2 (MISR, 30 Epochs) & $0.1183 \pm 0.0124$ \\
        Model 3 (MISR, 60 Epochs) & $\mathbf{0.1179 \pm 0.0129}$\\
        \hline       
    \end{tabular}
\end{center}
\end{table}

\section{Conclusion}

Separate inverse networks (and MISR) were used due to limited computational resources. Our model's performance could likely be improved by applying PCA with dimensionality reduction to the HSI scattering coefficients \cite{ScatGAN}, allowing us to train a single inverse network. This could also allow for deeper scattering transforms (i.e. with more details). We wil also explore further improvement through use of scattering transforms better suited to HSI data \cite{FourierScat}.





\label{sec:refs}

\bibliographystyle{IEEEbib}
\bibliography{refs}

\end{document}